\documentclass{article}

\usepackage{arxiv}

\usepackage[utf8]{inputenc} % allow utf-8 input
\usepackage[T1]{fontenc}    % use 8-bit T1 fonts
\usepackage{hyperref}       % hyperlinks
\usepackage{url}            % simple URL typesetting
\usepackage{booktabs}       % professional-quality tables
\usepackage{amsfonts}       % blackboard math symbols
\usepackage{nicefrac}       % compact symbols for 1/2, etc.
\usepackage{microtype}      % microtypography
\usepackage{lipsum}
\usepackage{graphicx}

\title{\texttt{\lowercase{cellanneal}}: A user-friendly deconvolution software for omics data}

\author{
Lisa Buchauer and Shalev Itzkovitz\\
  Department of Molecular Cell Biology, Weizmann Institute of Science, Rehovot, Israel\\
}

\begin{document}
\maketitle

\begin{abstract}
We introduce \texttt{cellanneal}, a python-based software for deconvolving bulk RNA sequencing data. \texttt{cellanneal} relies on the optimization of Spearman's rank correlation coefficient between experimental and computational mixture gene expression vectors using simulated annealing. \texttt{cellanneal} can be used as a python package or via a command line interface, but importantly also provides a simple graphical user interface which is distributed as a single executable file for user convenience. The python package is available at \url{https://github.com/LiBuchauer/cellanneal}, the graphical software can be downloaded at \url{http://shalevlab.weizmann.ac.il/resources}.
\end{abstract}

% keywords, separate by \and

\section{Introduction}\label{sec:intro}
Single-cell sequencing methods enable precise characterization of gene expression patterns in individual cells. However, they may provide inaccurate information about the cell type composition of samples, as required preprocessing procedures such as tissue dissociation or cell sorting affect viability of different cell types to varying extent \cite{erdmann2021likelihood}. Further, especially in the clinical context, single-cell sequencing of patient samples is currently not routinely applied because of high cost and required expertise, while bulk sequencing is more prevalent. 

For these reasons, computational deconvolution methods are gaining popularity in basic and clinical research. Computational deconvolution approaches infer the cell type proportions constituting a given bulk RNA sample based on separately obtained cell type reference data. Several computational deconvolution methods have been developed in the last decade and have contributed to our understanding of tissue composition \cite{cobos2020benchmarking, sturm2019benchmarking}. Generally, during deconvolution, the computational mixture is constructed from a set of cell type fractions and reference gene expression vectors for each of the participating cell types, most commonly derived from single-cell data. The cell type fractions are then iteratively changed until agreement between the \emph{in silico} gene expression vector and the observed bulk sample gene expression vector is optimal by a measure of choice. Here, published methods rely almost exclusively on minimizing the sum of squared residuals between bulk and computationally mixed vectors. Algorithms for such optimization problems are readily available and include variants of least squares regression (e.g. weighted least squares regression \cite{racle2017simultaneous}, non-negative least squares regression \cite{wang2019bulk, jew2020accurate} or least trimmed squares \cite{hao2019fast}) and support vector regression \cite{newman2015robust, newman2019determining}. 

However, least squares-based optimization is faced with a particular challenge in bulk RNAseq deconvolution because of the highly skewed nature of mRNA copy number distributions, ranging from less than 1 to more than 10,000 average mRNA copies per cell \cite{li2016comprehensive, schwanhausser2011global}. In such settings, optimization results may be strongly influenced by few highly expressed genes and are thus not robust to noise or platform effects influencing the readout of these genes. Identifying the right genes for deconvolution becomes a task in itself \cite{aliee2021autogenes}. As a result, deconvolution methods may yield inferred mixed gene expression vectors that do not correlate well with measured bulk gene expression. 

\section{Statement of Need}\label{sec:statement}
Making sense of bulk RNA sequencing datasets often requires analysis of the cell type composition of the samples. This is particularly relevant in clinical samples that analyze the transcriptome of tissues or tumors which consist of epithelial, stromal and immune cell types. In parallel, publicly available single-cell data sets enable precise characterization of the expression signature of multiple individual cell types. However, software tools for computational bulk deconvolution are often slow, non-robust and not easy to use. Some existing methods address the aspect of user-friendliness by providing graphical web interfaces, but submitting sensitive medical data to an external web server is not always compatible with privacy legislation.

To address these challenges, we have developed \texttt{cellanneal}, a deconvolution approach that uses Spearman’s rank correlation coefficient between synthetic and bulk gene expression vectors as the optimization procedure’s objective function. Because this correlation measure is calculated from ranks rather than absolute data values, each gene influences the optimization result to a similar extent. Users are encouraged to include as many informative genes as possible in the analysis. \texttt{cellanneal} optimizes cell type fractions by simulated annealing, a flexible, rapid and robust algorithm for global optimization \cite{kirkpatrick1983optimization, 2020SciPy-NMeth}. \texttt{cellanneal} can be used as a python package, via its command line interface or via a user-friendly graphical software which runs locally. Its typical processing time for one mixture sample is below one minute on a desktop machine.

\section{Availability and Features}\label{sec:feat}
The python package and command line interface are available at \url{https://github.com/LiBuchauer/cellanneal} and can be installed using \texttt{pip}. The graphical software for Microsoft Windows and MacOS can be downloaded at \url{http://shalevlab.weizmann.ac.il/resources} and does not require installation. Instructions for installation and use as well as general documentation is available at \url{https://github.com/LiBuchauer/cellanneal}.

The python package provides functions for the three main steps of a deconvolution analysis with \texttt{cellanneal}: identification of a gene set for deconvolution, deconvolution using simulated annealing, and plotting the results. A quick start workflow is available as part of the documentation. For the command line interface and the graphical user interface, these three steps are combined into one call (click).

\texttt{cellanneal} runs which were started from either the command line or the graphical user interface produce a collection of result files including tabular deconvolution results (cell type fractions for each sample) and figures illustrating these cell type distributions. Further, \texttt{cellanneal} computes and stores the gene-wise fold change between the observed bulk expression and the estimated expression based on the inferred cell type composition. This enables identifying genes for which expression may be specifically induced or inhibited in the bulk sample compared to the single cell reference. Such genes may be of biological or medical interest.

\texttt{cellanneal} relies on the python packages \texttt{scipy} \cite{2020SciPy-NMeth}, \texttt{numpy} \cite{harris2020array}, \texttt{pandas} \cite{reback2020pandas}, \texttt{seaborn} \cite{Waskom2021}, \texttt{matplotlib} \cite{Hunter:2007} and \texttt{tkinter} \cite{lundh1999introduction}.

\section{Acknowledgements}\label{sec:ackno}
We thank all members of the Itzkovitz lab for valuable feedback and for testing the software.

L.B. is supported by the European Molecular Biology Organization under EMBO Long-Term Fellowship ALTF 724-2019. S.I. is supported by the Wolfson Family Charitable Trust, the Edmond de Rothschild Foundations, the Fannie Sherr Fund, the Helen and Martin Kimmel Institute for Stem Cell Research grant, the Minerva Stiftung grant, the Israel Science Foundation grant No. 1486/16, the Broad Institute‐Israel Science Foundation grant No. 2615/18, the European Research Council (ERC) under the European Union’s Horizon 2020 research and innovation program grant No. 768956, the Chan Zuckerberg Initiative grant No. CZF2019‐002434, the Bert L. and N. Kuggie Vallee Foundation and the Howard Hughes Medical Institute (HHMI) international research scholar award. 

%\begin{figure}
%  \centering
%  \includegraphics{Figure1_20190326}
%  \caption{lalacap}
%  \label{fig:fig1}
%\end{figure}

\bibliographystyle{unsrt}  
\bibliography{references}

\end{document}